\documentclass[aps,physrev,reprint,superscriptaddress]{revtex4-2}

\usepackage{graphicx}
\usepackage{subcaption}
\usepackage{amsmath}
\usepackage{amssymb}

\usepackage{hyperref}

\begin{document}


\title{Nanoparticle Networks for Neuromorphic Computing}


\author{Jonas Mensing}
\affiliation{Theory of Complex Systems Group, Institute of Physical Chemistry, University of Münster, Münster, Germany}
\author{Wilfred G. van der Wiel}
\affiliation{NanoElectronics Group, MESA+ Institute for Nanotechnology, Center for Brain-Inspired Nano Systems (BRAINS), University of Twente, Enschede, Netherlands}
\affiliation{Institute of Physics, University of Münster, Münster, Germany}
\author{Andreas Heuer}
\email{Contact author: andheuer@uni-muenster.de}
\affiliation{Theory of Complex Systems Group, Institute of Physical Chemistry, University of Münster, Münster, Germany}

\date{\today}

\begin{abstract}
\centering
Physical computing leverages complex dynamical systems for energy-efficient data processing. In this work, we present a neuromorphic architecture based on metallic nanoparticles interconnected by molecular junctions on a $\text{SiO}_2$/Si substrate. We demonstrate that surrounding static control electrodes transform this nanoparticle network from a passive reservoir into a tunable nonlinear dynamical system. By analyzing how these electrodes route simple one-dimensional voltage inputs into multidimensional signal responses, we establish three core design rules to maximize computational performance. First, operating near the system's cutoff frequency achieves an optimal balance between nonlinear charge tunneling and linear capacitive memory. Second, tuning the underlying $\text{SiO}_2$ thickness sets the electrostatic screening length and dictates the memory type. Thick oxide layers reduce the screening length, causing networks larger than this length to transition into a persistent, non-volatile-like regime. Conversely, networks smaller than the screening length exhibit only fading memory. Third, introducing structural disorder via heterogeneous molecular junctions overcomes inherent limits on expressivity. While a network's computational expressivity scales with its physical size, it is ultimately capped by the screening length. Breaking internal spatial symmetries with localized disorder bypasses this saturation, allowing control voltages to independently manipulate specific signal amplitudes and phases, universally maximizing performance for dynamic neuromorphic applications.
\end{abstract}


\maketitle

\section{\label{sec:introduction}Introduction}

The progression of data-driven algorithms and artificial intelligence has revealed the fundamental limitations of the von Neumann computing architecture. Shuttling data between physically separated processing and memory units are accompanied with escalating energy costs.~\cite{garcia2019estimation, luccioni2020estimating, faiz2024llmcarbon, ahrabi2025ai} This so-called von Neumann bottleneck gives rise to a paradigm shift towards alternative neuromorphic computing architectures. Neuromorphic computing is characterized by energy efficient, parallel, spatiotemporal processing capabilities which are also intrinsic to biological neural networks.~\cite{sangwan2020neuromorphic, enuganti2025neuromorphic, ganguly2019towards} These computing frameworks use nonlinear, high-dimensional, and time-dependent dynamics of nanomaterials to perform computations on hardware.

Within this domain, physical Reservoir Computing (RC) treats a complex, recurrently connected dynamical system as an untrained black box or reservoir.~\cite{tanaka2019recent, qi2023physical} The reservoir's purpose is to project low-dimensional input signals into a high-dimensional feature space through nonlinear transformations~\cite{sillin2013theoretical,dang2024effect,srikimkaew2023high}. However, RC relies on a linear readout layer to be trained via regression to extract the desired target function. Even though there are online approaches for implementing the readout training on hardware~\cite{ma2023integrated,smerieri2012analog,zhong2022memristor,han2025reconfigurable}, most RC approaches still rely on a slow and digital readout phase during post-processing on standard computing hardware.

In this work, we study nanoparticle networks for neuromorphic computing applications.~\cite{bose2015evolution,mensing2024kinetic} Self-assembled networks of metallic nanoparticles interconnected by insulating organic molecules provide a designless RC substrate. The interplay of capacitive interactions and stochastic single-electron tunneling spans a collection of accessible dynamical states. In contrast to conventional RC, where the physical substrate is fixed and only a readout layer is trained, we bypass the readout layer entirely. Here, the physical device itself is tuned by static control voltages. The network is surrounded by a set of electrodes, of which we use one to voltage-encode one-dimensional input data and one grounded electrode to measure the network's electric current response. The remaining electrodes serve as these static controls. By manipulating the inner capacitive and charge tunneling dynamics, the controls reshape the internal dynamical response before readout, allowing the same network to realize different nonlinear transformations. Accordingly, these controls only need to be configured once to find a set of voltages matching the desired input-output functionality.

To establish the physical foundation for this architecture, we conduct a theoretical and numerical analysis of the network's electrostatic and dynamical properties. We first investigate the capacitive interaction profile of the network, demonstrating how experimental design parameters govern the structural complexity of the state space required for neuromorphic computation.~\cite{enguita2023principal, he2025role, udaya2023optimization}

We then study the uncontrolled system to map the distinct operational regimes driven by nonlinear charge tunneling and linear capacitive displacement. By evaluating the frequency-dependent response, we identify a fundamental bandwidth limit that establishes the optimal cutoff frequency required to balance phase-delay memory with high-order harmonic generation, a feature that is particularly critical for neuromorphic applications.~\cite{petrauskas2022nonlinear,viero2018light,sillin2013theoretical,cohen2012second} We also demonstrate that the network's tunable memory dynamics are governed by the distance of each nanoparticle to the substrate, which dictates the electrostatic screening length. While fully isolated systems possess an infinite screening length and exhibit fading memory, screened networks transition into a persistent, non-volatile regime when the network length exceeds this screening length. The emergence of such persistent memory states is vital for long-term weight storage.~\cite{burr2017neuromorphic, chakraborty2020pathways,carroll2022optimizing,oh2025coexistence, song2025self}

Finally, we quantify the network's true high-dimensional expressivity under active electrode control. Here, we utilize a phase-aware effective volume as a singular metric across a large ensemble of variable control drives. Using this singular metric allows us to capture the full scope of device tuneability, rather than relying on statistical distributions of isolated, one-dimensional projections.~\cite{bose2015evolution,mensing2024kinetic,tertilt2024critical} Through this metric, we reveal a fundamental topological limit. Increasing the system size expands the effective volume, but this expansion is bounded by the screening length. Furthermore, in a highly symmetric network, different control configurations often produce redundant transformations, which inherently bottlenecks the system's expressivity. To push past these limitations, we introduce structural disorder via heterogeneous molecular junctions. Breaking the internal spatial symmetries allows the control electrodes to address more independent dynamical modes. Consequently, the static control voltages can independently manipulate specific harmonic amplitudes and their relative phases, universally expanding the high-dimensional response space.

\section{\label{sec:theory_methods}Theory and Methods}

\subsection{\label{sec:device_setup}Device Architecture and Experimental Setup}

\begin{figure}[t]
    \centering
    \includegraphics[width=0.48\textwidth]{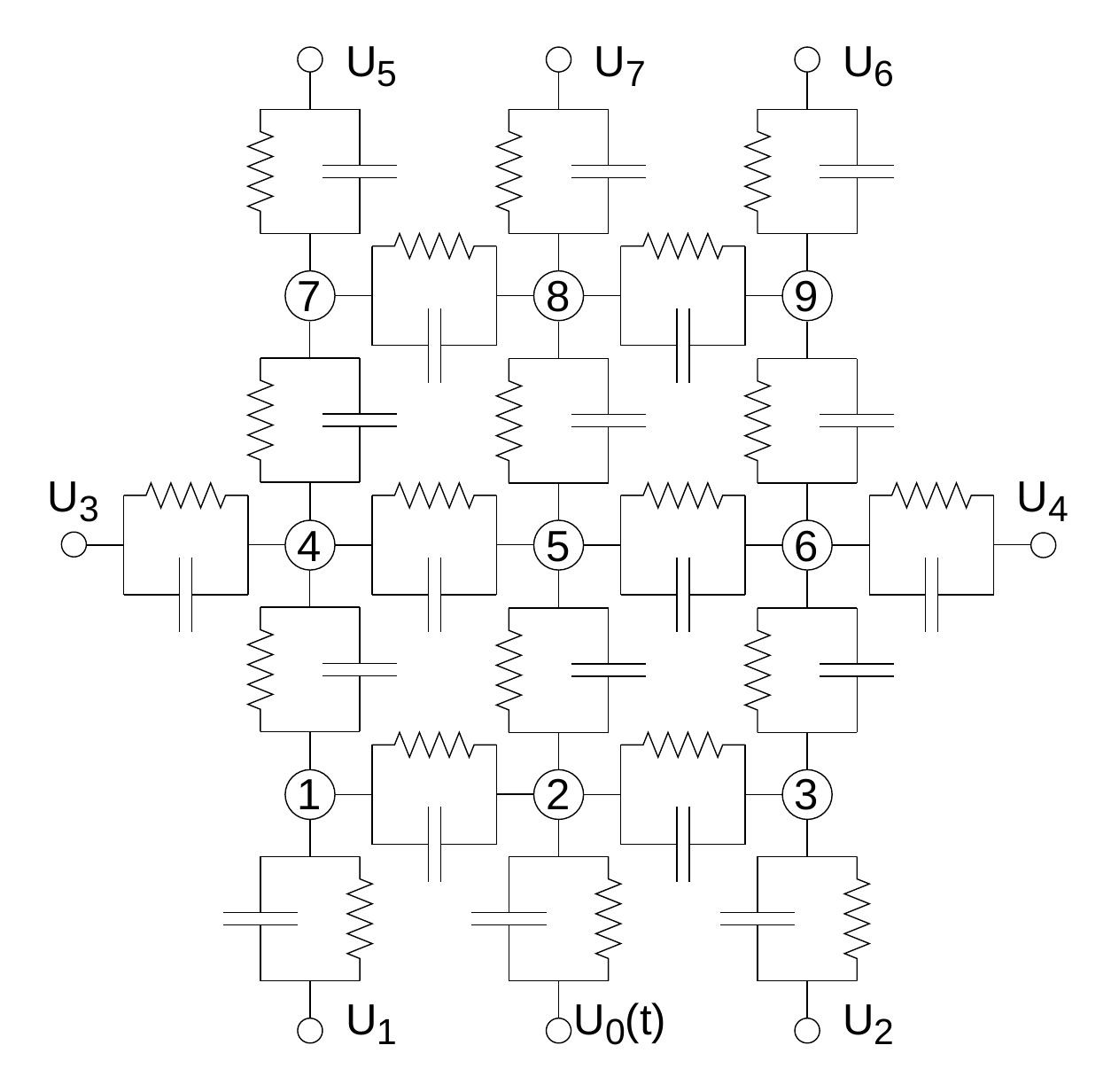}
    \caption{\textbf{Circuit Schematic of a $3 \times 3$ Nanoparticle Network interfaced with 8 Electrodes.} Charge transport between adjacent islands $i$ and $j$ is governed by a parallel combination of a nonlinear tunnel resistance $R_{ij}$ and a junction capacitance $C_{ij}$. Furthermore, the oxide layer establishes a local substrate coupling capacitance $C_i$ at each individual nanoparticle, acting as a distributed ground. Electrode voltages are denoted $U_i$.}
    \label{fig:device_schematic}
\end{figure}

We study two-dimensional square lattices of metallic nanoparticles defined by their edge length $L$ with $N_{P} = L \times L$ nanoparticles in total. Throughout this work, the lattice size $L$ will be varied from $L=3$ ($N_P = 9$) to $L=15$ ($N_P = 225$), which allows us to investigate system size effects. Furthermore, we assume that the network is deposited on a $\text{SiO}_2$ dielectric layer of thickness $d_\textrm{ox}$ above a highly doped Si substrate. The resulting network is coupled to the external environment via $N_e = 8$ electrodes positioned symmetrically around the lattice perimeter. We classify the role of each electrode as follows:
\begin{itemize}
    \item \textbf{Input Electrode:} A single electrode ($U_0$, bottom-center) acts as the dynamic driving source. Depending on the experiment, it injects either a sinusoidal AC voltage $U_0(t) = \tilde{U}_0 \sin(2\pi f_0 t)$ or a discrete step-function $U_0(t) = U_0$
    \item \textbf{Output Electrode:} The temporal response of the network is read out at the drain electrode ($U_7$, top-center), which is clamped to ground ($U_7(t) = 0~\mathrm{V}$). The measured observable is the total time-dependent output current $I(t)$, which is the sum of both the current based on nonlinear resistive tunneling transport $I^\textrm{tun}(t)$ and the displacement current $I^\textrm{disp}(t)$ resulting from capacitive interactions.
    \item \textbf{Control Electrodes:} The remaining six electrodes are set to static voltages, $\vec{U}_{\text{ctrl}} = \{U_{1}, \dots, U_{6}\}$. These controls serve as a tuning knob for the output electrode electric current $I(t)$ to input electrode voltage $U_0(t)$ dependence.
\end{itemize}
FIG.~\ref{fig:device_schematic} shows a circuit schematic of a $L = 3$ nanoparticle network excluding the electrostatic coupling to the substrate. Each nanoparticle is electrostatically coupled (capacitor) and tunnel-coupled (resistor) to a neighboring nanoparticle or electrode.

\subsection{\label{sec:electrostatic}Electrostatics and Network Dimensionality}


\begin{figure}[tb]
    \centering
    \includegraphics[width=0.48\textwidth]{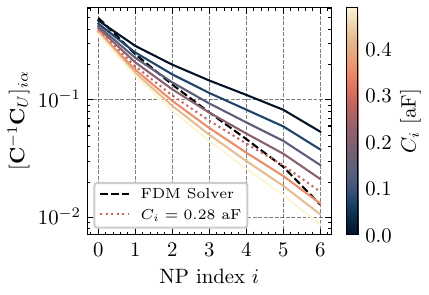}
    \caption{\textbf{Spatial Screening.} Spatial decay of the electrostatic coupling strength $[\mathbf{C}^{-1}\mathbf{C}_U]_{i\alpha}$ from the input electrode at position $i=0$ into the network bulk. The exact 3D FDM solution (dashed black line) demonstrates exponential screening. The approximation at variable values of $C_i$ replicates this exponential decay at different slopes.}
    \label{fig:spatial_screening}
\end{figure}

The network's electrostatic interactions are dictated by the $N_p \times N_p$ capacitance matrix $\mathbf{C}$, which relates the vector of total nanoparticle charges $\vec{q}(t)$ to the vector of nanoparticle potentials $\vec{\phi}(t)$. The diagonal elements represent the total capacitance of nanoparticle $i$ to the surrounding electrostatic environment $[\mathbf{C}]_{ii} = C_i + \sum_{j \neq i} C_{ij}$, while the off-diagonal elements $[\mathbf{C}]_{ij} = -C_{ij}$ represent the mutual capacitances between nanoparticles $i$ and $j$. Here, $C_i$ is the coupling of nanoparticle $i$ to the underlying grounded Si substrate. The electrostatic coupling of electrode voltages $\vec{U}$ to nanoparticle potentials is summarized in the $N_p \times N_e$ electrode-to-nanoparticle capacitance matrix $\mathbf{C}_U$ containing the positive mutual capacitance values $C_{i\alpha}$ between nanoparticle $i$ and electrode $\alpha$. 

The inverse of the capacitance matrix $\mathbf{C}^{-1}$ dictates the total electrostatic energy of the network
\begin{equation}\label{eq:e_energy}
    E_C = \frac{1}{2}\vec{q}^\top\mathbf{C}^{-1}\vec{q}
\end{equation}
and the single-electron charging energy of nanoparticle $i$
\begin{equation}\label{eq:e_energy_single}
    E_{C,i} = \frac{1}{2}e^2[\mathbf{C}^{-1}]_{ii}.
\end{equation}
FIG.~\ref{app:charging_energies} proofs the threshold voltage $E_{C,i}/e$ associated to single-electron charging energy to increase with lattice size $L$. However, this increase is only unbounded when nanoparticles do not couple to a beneath substrate, i.e. $C_i = 0$. In the case of $C_i = 0.28~\mathrm{aF}$, the voltage values approach the same distribution for $L \to \infty$. This marks two important consequences already. For a fair theoretical comparison between different network sizes $L$ or substrate couplings $C_i$, we have to scale the electrode voltage vector $\vec{U}$ according to this voltage threshold $E_{C,i}/e$. Here we choose the maximum voltage threshold $\max_i\{E_{C,i}/e\}$ summarized in TABLE~\ref{tab:voltage_scaling} as our reference voltage for each network. Secondly, when designing experiments, choosing a smaller oxide thickness will bound the operational voltage regime allowing to ignore the exact number of nanoparticles at least in case the network is chosen to be large enough.

The total charge $\vec{q}(t)$ on the nanoparticles is a superposition of the discrete integer excess charges $\vec{q}_{\text{int}}$ (resulting from stochastic charge tunneling events) and the continuous induced charges by polarization driven by the time-dependent external electrode voltages $\vec{U}(t)$:
\begin{equation}\label{eq:total_charge}
    \vec{q}(t) = \vec{q}_{int} + \mathbf{C}_U \vec{U}(t).
\end{equation}
Consequently, the potential vector $\vec{\phi}(t)$ is updated by both discrete stochastic charge tunneling events and the deterministic continuous input drive: 
\begin{equation}\label{eq:pot_vec}
    \vec{\phi}(t) = \mathbf{C}^{-1} \vec{q}(t) = \mathbf{C}^{-1} (\vec{q}_{int} + \mathbf{C}_U \vec{U}(t)).
\end{equation}

The capacitance values composing the capacitance matrices $\mathbf{C}$ and $\mathbf{C}_U$ are analytically unknown. Given the many-body system, capacitance values between two conductive elements have to consider higher-order screening effects. To establish a correct electrostatic baseline we first establish a hierarchy of capacitive interaction using a 3D Finite Difference Method (FDM). We use geometric parameters consistent with the experimental device of \cite{bose2015evolution} (oxide thickness of $35~\mathrm{nm}$, inter-particle spacing of $1~\mathrm{nm}$ and nanoparticle radii of $10~\mathrm{nm}$) and solve the Poisson equation to derive the exact capacitance matrices $\mathbf{C}$ and $\mathbf{C}_U$ for a representative $7 \times 7$ lattice. The resulting matrix in FIG.~\ref{app:capacitance_matrix_FMD} confirms that electrostatic interactions are only present between nearest neighbors given the dense network packing and next neighbor interactions being screened by nearest neighbor conductors. Also, the hierarchy in average capacitance values, shown in FIG.~\ref{app:capacitance_comparsion}, state that the network is dominated by nanoparticle-nanoparticle or electrode-nanoparticle interactions ($\bar{C}_{ij} \approx \bar{C}_{i\alpha} \approx 2.0~\mathrm{aF}$) compared to the coupling of each nanoparticle $i$ to the beneath substrate ($\bar{C}_i \approx 0.5~\mathrm{aF}$) which is dictated by the chosen oxide thickness of $d_\textrm{ox} = 35~\mathrm{nm}$. 

To maintain computational scalability for the upcoming network size sweeps without requiring 3D FDM calculations for every system size $L$, we employ an approximation for the remainder of this work. The FDM benchmark demonstrates that the mutual capacitance between nearest-neighbor nanoparticles is well approximated by the analytical two-body solution for isolated spherical conductors of radii $r_i$ and $r_j$ separated by a center-to-center distance $d_{ij}$:~\cite{lekner2011capacitance}
\begin{equation}\label{eq:mutual_cap}
    C_{ij} = 4 \pi \epsilon_0 \epsilon_r \frac{r_i r_j}{d_{ij}} \sinh(\zeta_{ij})\sum_{n=1}^\infty\frac{1}{\sinh(n\zeta_{ij})}
\end{equation}
where $\cosh(\zeta_{ij}) = (d_{ij}^2-r_i^2-r_j^2)/(2r_i r_j)$ and $\epsilon_r = 2.6$ is the relative permittivity of the insulating organic molecule 1-octanethiol. For the parameters we put into the FDM solver i.e. nanoparticles of radius $r_i = r_j = 10~\mathrm{nm}$ at a center-to-center distance $d_{ij} = 21~\mathrm{nm}$, we get a mutual coupling of $C_{ij} = 2.28~\mathrm{aF}$ which is close to the average value derived from the FDM solver $\bar{C}_{ij} \approx 2.0~\mathrm{aF}$. This analytical approximation is further justified when admitting the structural complexity in real-world self-assembly, where especially inter-particle spacing but also effective local dielectric constants are unknown leading to unavoidable discrepancies between real-world devices and the FDM approach. Therefore, sticking to equation~\eqref{eq:mutual_cap} to approximate the particle-particle and electrode-particle interactions ensures a phenomenological agreement with experimental values. Given the fact, we also investigate the impact of the oxide thickness $d_\text{ox}$ in this work, we set the related nanoparticle to substrate coupling to either $C_i = 0.28~\mathrm{aF}$, $C_i = 0.07~\mathrm{aF}$, or $C_i = 0~\mathrm{aF}$. These three values are all below the FDM estimated average coupling of $\bar{C}_i \approx 0.5~\mathrm{aF}$ ensuring a mutual interaction dominant device. The distance to which a nanoparticle charge induces potentials on neighboring nanoparticles is defined by the electrostatic screening length $\Lambda$. Using a 1D continuum approximation (equation~\eqref{app:c_lap}) for the screening length (equation~\eqref{app:soliton_Length}) shows that the ratio of inter-nanoparticle capacitance to the nanoparticle to substrate capacitance dictates this electrostatic cross-talk via $\Lambda = \sqrt{C_{ij} / C_i}$. Plugging in our specific values for $C_{ij}$ and $C_i$, we find that the electrostatic interaction decay over characteristic lengths of $\Lambda \approx 3$, $\Lambda \approx 6$, and $\Lambda \to \infty$ nanoparticles, marking distinct operational regimes.

The element $M_{i\alpha}$ of the electrode-to-network coupling matrix $\mathbf{M} = \mathbf{C}^{-1}\mathbf{C}_U$ (with shape $N_p \times N_e$, defined in equation.~\eqref{eq:pot_vec}) dictates how the voltage of electrode $\alpha$ capacitively couples to nanoparticle $i$. Plotting the elements $M_{i\alpha}$ along a path from the input to the output electrode reveals an exponential spatial screening profile (FIG.~\ref{fig:spatial_screening}), with a decay rate defined by the screening length $\Lambda$ and thus by the substrate coupling $C_i$. Furthermore, this electrode-to-network coupling matrix maintains a non-vanishing singular-value spectrum across all considered substrate couplings, confirming that the control electrodes address independent directions in the nanoparticle-potential space (FIG.~\ref{fig:control_dimension}).

\subsection{\label{sec:kmc_model}Stochastic Charge Transport}

The time evolution of the network's state is simulated using a Kinetic Monte Carlo (KMC) algorithm based on the orthodox theory of single-electron tunneling.~\cite{likharev2002single,mensing2024kinetic,wasshuber2001computational} We restrict our model to sequential first-order tunneling, assuming that higher-order co-tunneling processes are negligible.

The probability per unit time of a tunneling event from nanoparticle $i$ to an adjacent nanoparticle $j$ is governed by the corresponding change in the system's free energy $\Delta F_{i \to j}(t) = \Delta E_C - \Delta W(t)$. This energy change is dictated by the potential equation~\eqref{eq:pot_vec}:
\begin{align}\label{eq:free_energy}
    \Delta F_{i \to j}(t) = &-e [\phi_i(t) - \phi_j(t)] \nonumber \\
    &+ \frac{e^2}{2} \left( [\mathbf{C}^{-1}]_{ii} + [\mathbf{C}^{-1}]_{jj} - 2[\mathbf{C}^{-1}]_{ij} \right)
\end{align}
where the first term represents the time-dependent electrostatic work done by the local potential difference, and the second term summarizes the constant charging energy cost required to overcome the Coulomb blockade. The transition rate $\Gamma_{i \to j}(t)$ for this specific tunneling event is determined by the rate equation:
\begin{equation}\label{eq:tunneling}
    \Gamma_{i \to j}(t) = \frac{1}{e^2 R_{ij}} \frac{\Delta F_{i \to j}(t)}{\exp[\Delta F_{i \to j}(t) / k_B T] - 1}
\end{equation}
where $R_{ij}$ is the tunnel resistance of the junction between nodes $i$ and $j$, and $T$ is the network's temperature. Throughout this work, the temperature will be set to $T = 0.1~\mathrm{K}$, ensuring the suppression of thermally activated tunneling (see section~\ref{app:temperature_limit}).

To investigate the role of structural disorder in section~\ref{sec:disorder}, we define the tunnel resistance $R_{ij}$ of each molecular junction as a random variable drawn from a binary distribution:
\begin{equation}\label{eq:bimonal_dis_dist}
    R_{ij} = 
    \begin{cases} 
      R_1 = 25 \, \mathrm{M\Omega} & \text{with probability } p = 2/3 \\
      R_2 = \kappa R_1 & \text{with probability } 1-p = 1/3 
   \end{cases}
\end{equation}
This mathematically mimics the experimental capability to synthesize networks with varying molecular junctions. The disorder strength is parameterized by the resistance contrast ratio $\kappa > 1$. The probability of forming a highly conductive $R_1$ junction is fixed at $p = 2/3$. For an infinite 2D square lattice, the analytical bond percolation threshold is $p_c = 1/2$.~\cite{yonezawa1989percolation} By fixing $p > p_c$, the local spatial symmetry is disrupted, while still a connected, percolating cluster of high-conductance pathways persists from the input to the output electrode across the majority of spatial realizations. For statistical robustness, we simulate an ensemble of 32 individually seeded random spatial realizations for a given value of $\kappa = 4$. The kinetic Monte Carlo model is outlined in section~\ref{app:kmc}. 

\begin{figure}[bt]
    \centering
    \includegraphics[width=0.48\textwidth]{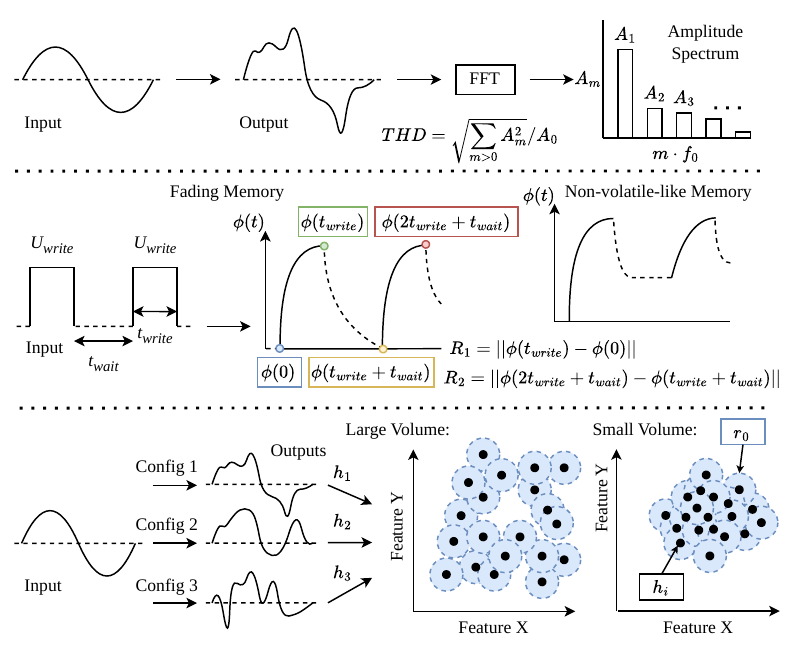}
    \caption{\textbf{Top (Characterizing Nonlinearity):} Baseline nonlinearity is isolated using a sinusoidal input with grounded controls. The Total Harmonic Distortion (THD) is calculated from the FFT amplitude spectrum as the ratio of higher-order harmonics to the fundamental amplitude. \textbf{Middle (Characterizing Memory):} Temporal memory is probed via two identical input pulses separated by a wait time $t_\mathrm{wait}$. The qualitative response curves for the potential vector $\phi(t)$ contrast a fading memory that relaxes toward its initial state with a non-volatile-like memory that retains its polarization. Memory capacity is quantified by the state displacements during the first ($R_1$) and second ($R_2$) pulse. \textbf{Bottom (Characterizing Tuneability):} To measure expressivity, varying static controls generate distinct current responses mapped to harmonic feature vectors ($h_i$). The reachable state space is evaluated by its effective volume, conceptually contrasting a highly tuneable system (large volume of widespread, distinct states) against a poorly tuneable system (small volume of clustered states overlapping within radius $r_0$).}
    \label{fig:metrics}
\end{figure}

To characterize the computational properties of the nanoparticle network, we utilize metrics for its nonlinearity, memory, and control-driven expressivity (see FIG.~\ref{fig:metrics}). Crucially, these metrics evaluate the network under two different operating conditions. While the nonlinearity and memory metrics isolate the network's baseline response by clamping all control electrodes to ground, the expressivity analysis actively varies these control voltages to map out the reachable dynamical state space.

\subsection{Defining Nonlinearity and Memory}

\paragraph{Total Harmonic Distortion:} First we evaluate the baseline nonlinearity of the network and apply a continuous sinusoidal drive to the input electrode, given by $U_0(t) = \tilde{U}_0 \sin(2\pi f_0 t)$. The resulting time-domain output current response $I(t)$, deviates from a perfect sinusoid due to the nonlinear nature of single-electron tunneling across the Coulomb blockade thresholds. To quantify the magnitude of this nonlinear transformation independently of any phase behavior, we perform a Fast Fourier Transform (FFT) on the steady-state current $I(t)$ to extract the amplitude spectrum $A_m$ for each $m$-th harmonic frequency ($m \cdot f_0$). The proportion of total signal power shifted into these higher-order harmonics relative to the linear fundamental response ($A_1$) is defined by the Total Harmonic Distortion (THD):
\begin{equation}\label{eq:THD}
    \text{THD} = \frac{\sqrt{\sum_{m > 1} A_m^2}}{A_1}.
\end{equation} 

\paragraph{Paired-Pulse Ratio:} While THD characterizes the power of the network's continuous nonlinearity, it does not capture its temporal memory capacity. To probe this, we implement a paired-pulse experiment.~\cite{lee2020inverse, li2019fully} Instead of a harmonic drive, the input electrode is defined by a sequence of two identical, non-overlapping square voltage pulses of amplitude $U_\textrm{write}$ and duration $t_\textrm{write}$, separated by a zero-voltage wait time $t_\textrm{wait}$:
\begin{equation}
    U_0(t)=
    \begin{cases}
    U_{\mathrm{write}}, & 0 \le t < t_{\mathrm{write}},\\[4pt]
    0, & t_{\mathrm{write}} \le t < t_{\mathrm{write}}+t_{\mathrm{wait}},\\[4pt]
    U_{\mathrm{write}}, &
    t_{\mathrm{write}}+t_{\mathrm{wait}}
    \le t <
    2t_{\mathrm{write}}+t_{\mathrm{wait}}.
    \end{cases}
\end{equation}
The network's response to each pulse is defined as the Euclidean distance between its potential vector $\vec{\phi}(t)$ at the start and end of the respective pulse duration:
\begin{align}
R_1 &= \left\|\vec{\phi}(0) - \vec{\phi}(t_{\mathrm{write}}) \right\|_2, \nonumber \\
R_2 &= \left\|\vec{\phi}(t_{\mathrm{write}}+t_{\mathrm{wait}}) - \vec{\phi}(2t_{\mathrm{write}} + t_{\mathrm{wait}}) \right\|_2.
\end{align}
The network's memory is then quantified by the Paired Pulse Ratio (PPR):
\begin{equation}\label{eq:PPR}
    \text{PPR} = \frac{R_2 - R_1}{R_2 + R_1}.
\end{equation}
A value of $\text{PPR} \to -1$ indicates perfect state retention, meaning the network is still fully polarized by the first pulse, when the second pulse arrives. Conversely, $\text{PPR} \to 0$ signifies a complete loss of memory, where the network has entirely relaxed and responds to the second pulse as if it were an unperturbed system. By systematically sweeping $t_\textrm{wait}$ logarithmically, we map the network's characteristic forgetting curve. The write pulse duration is fixed to a sufficiently long value of $t_{\mathrm{write}} = 102.4~\mathrm{ns}$ to guarantee that the system reaches a fully polarized steady state before the pulse ends.

\begin{figure*}[th]
\centering
\begin{subfigure}[b]{0.5\textwidth}
\centering
\includegraphics[height=0.18\textheight]{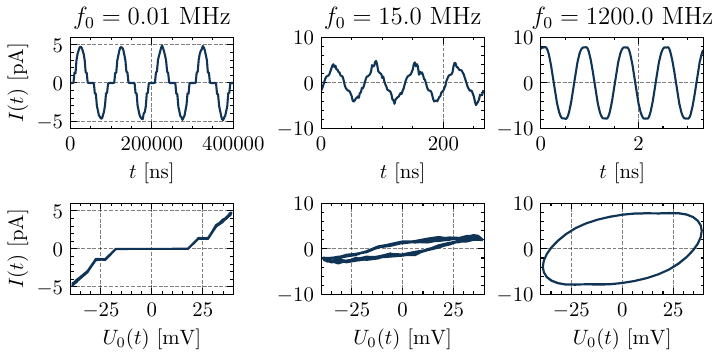}
\caption{Time Domain Waveforms}
\label{fig:time_domain_plots}
\end{subfigure}
\hfill
\begin{subfigure}[b]{0.46\textwidth}
\centering
\includegraphics[height=0.18\textheight]{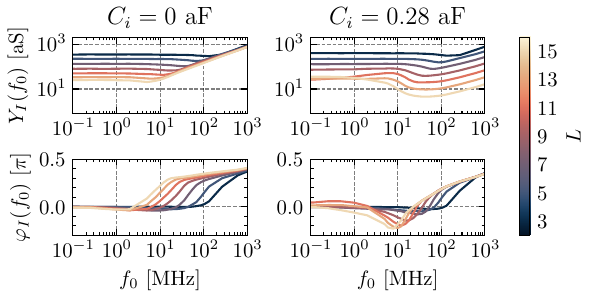}
\caption{Admittance and Phase Shift}
\label{fig:bode_plots}
\end{subfigure}
\caption{\textbf{Signal Processing Characteristics} \textbf{(a)} Time-domain and phase space plots of the total output current $I(t)$ given a sinusoidal input voltage $U_0(t) = \tilde{U}_0 \sin(2 \pi f_0 t)$ for a network of size $L=9$ and substrate coupling $C_i = 0.28~\mathrm{aF}$. These plots demonstrate the transition from an in-phase current $I(t) \approx I^\text{tun}$ dictated by nonlinear tunneling at small frequencies ($0.01~\mathrm{MHz}$), to a linear $\pi / 2$ phase-leading current $I(t) \approx I^\text{disp}$ defined by capacitive charge displacement ($1200~\mathrm{MHz}$). At intermediate frequencies ($15.0~\mathrm{MHz}$) both current types contribute to the total current $I(t) = I^\text{tun} + I^\text{disp}$ \textbf{(b)} Admittance $Y_I$ and phase shift $\varphi_I$ of the total output current $I(t)$. The total response transitions from a low-frequency, quasi-equilibrium resistive state to a high-frequency, capacitive shunt. In screened networks at substrate coupling $C_i = 0.28~\mathrm{aF}$, the RC network introduces a negative phase delay ($\varphi_I < 0$).}
\label{fig:time_domain_and_bode}
\end{figure*}

\paragraph{Effective Volume:} To evaluate the network's computational expressivity, we actively vary the static control voltages to sample $N_\textrm{cfg} = 1000$ distinct configurations $\vec{U}$. At each configuration $i$, the resulting steady-state current response to the sinusoidal drive is mapped into a phase-aware harmonic feature vector:
\begin{equation}\label{eq:feature_vector}
    \vec{h}_i = \left(\dots, \frac{A_m}{A_1} \cos(\Delta \phi_m), \frac{A_m}{A_1} \sin(\Delta \phi_m), \dots \right).
\end{equation}
Here, $\Delta \phi_m = \phi_m - m\phi_1$ is the relative phase of the $m$-th harmonic. By normalizing to the fundamental amplitude $A_1$ and explicitly omitting the $m=1$ order, $\vec{h}_i$ isolates the control-induced signal distortion within a $2(M-1)$-dimensional state space. The $N_\textrm{cfg}$ configurations for a specific network (length $L$, coupling $C_i$) form a local point cloud $\mathcal{H}_{L,C_i}$. To enable quantitative comparisons across different networks, all local clouds must be evaluated against a shared spatial scale. Therefore, we aggregate them into a single global point cloud, $\mathcal{H} = \bigcup \mathcal{H}_{L,C_i}$, and estimate the effective volume ($\mathcal{V}_\textrm{eff}$) using Monte Carlo integration:
\begin{enumerate}
    \item \textbf{Shared Radius ($r_0$):} We establish a universal spatial resolution based on the median inter-point distance of the entire global cloud:
    \begin{equation}\label{eq:v_eff_resolution}
        r_0 = \gamma \cdot \text{median}\left( ||\vec{h}_\mu - \vec{h}_\nu|| \right) \quad \text{for} \quad \vec{h}_\mu, \vec{h}_\nu \in \mathcal{H}.
    \end{equation}
    \item \textbf{Global Bounding Box:} We construct a single hyper-rectangular bounding box (volume $\mathcal{V}_{\text{box}}$) enclosing $\mathcal{H}$, and generate $N_{\text{MC}}$ uniformly distributed random sample points $\vec{s}_k$ within it.
    \item \textbf{Nearest Neighbor Hit:} A random sample $\vec{s}_k$ is registered as a success for a specific network if it falls within the radius $r_0$ of any data point in that network's local cloud:
    \begin{equation}
        \min_{\vec{h}_i \in \mathcal{H}_{L,C_i}} ||\vec{s}_k - \vec{h}_i|| \le r_0.
    \end{equation}
    \item \textbf{Volume Calculation:} The effective dynamic volume for the specific network is the fraction of its local hits scaled by the global bounding box volume:
    \begin{equation}\label{eq:eff_volume}
        \mathcal{V}_{\text{eff}} \approx \mathcal{V}_{\text{box}} \cdot \frac{N_{\text{hits}}}{N_{\text{MC}}}.
    \end{equation}
\end{enumerate}
Intuitively, $\mathcal{V}_{\text{eff}}$ measures how much of the harmonic response space can be reached by tuning the static control electrodes. A larger $\mathcal{V}_{\text{eff}}$ means that the same physical network can generate a broader set of distinguishable nonlinear dynamical responses.

\section{\label{sec:ac_response_bandwidth}Signal Processing}

To establish the fundamental signal processing regimes of different sized nanoparticle networks with lattice length $L$, we first evaluate the electric current response $I(t)$ at the grounded output electrode to a sinusoidal input electrode signal $U_0(t) = \tilde{U}_0\sin(2 \pi f_0 t)$. The remaining control electrodes are grounded. The driving amplitude $\tilde{U}_0$ is defined as described in section~\ref{sec:electrostatic}.

FIG.~\ref{fig:time_domain_plots} shows exemplary time-domain and phase space curves of $I(t)$ for a $9 \times 9$ lattice. FIG.~\ref{fig:bode_plots} presents the total admittance $Y_I(f_0) = \sigma_I(f_0) / \tilde{U}_0$, $\sigma_I(f_0)$ is the standard deviation of the electric current. It also displays the phase shift, $\varphi_I(f_0)$ of the output's fundamental harmonic relative to the input signal. By isolating the fundamental frequency, we evaluate the system's linear signal processing characteristics and phase-shifting behavior despite the presence of nonlinear distortion. At small frequencies ($f_0 < 1~\mathrm{MHz}$) only resistive tunnelling currents (equation~\eqref{eq:I_tun}) contribute to $I(t) \approx I^\text{tun}$ and the network is in a quasi-equilibrium state at zero phase shift. Here, the period of $U_0(t)$ exceeds the tunnelling time $\Delta t$ of equation~\eqref{eq:dwell_time}. Computationally, the network acts as a static, memoryless feedforward layer where $Y_I$ is flat and defined strictly by the series resistance of the junctions. At large frequencies ($f_0 > 100~\mathrm{MHz}$), the networks transition into a regime of capacitive shunting with only displacement currents defined in equation~\eqref{eq:I_disp} contributing i.e. $I(t) \approx I^\text{disp}$. The total admittance rises continuously, and the phase approaches the maximum capacitive $0.5\pi$ lead. In this regime, the physical substrate acts as a high-pass filter responding primarily to the rate of change of the input signal rather than its absolute magnitude. In the transition region ($1~\mathrm{MHz} \le f_0 \le 100~\mathrm{MHz}$), both resistive and displacement components contribute to $I(t) = I^\text{tun} + I^\text{disp}$.

\begin{figure}[tbh]
    \centering
    \includegraphics[width=0.48\textwidth]{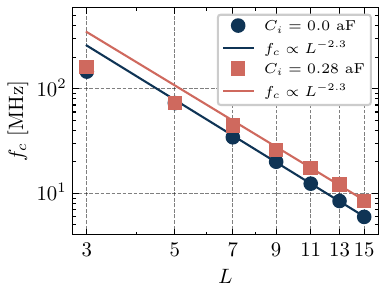}
    \caption{\textbf{Cutoff Frequency.} The cutoff frequency, marking the crossover at which tunnelling $I^\textrm{tun}$ and displacement $I^\textrm{dips}$ currents contribute equally to the total current $I(t)$ is independent of the substrate coupling $C_i$ i.e. oxide thickness $d_\textrm{ox}$. The network's bandwidth is size dependent and dictated by a scaling of $f_c \propto L^{-2.3}$.}
    \label{fig:bandwidth}
\end{figure}

\begin{figure}[b]
    \centering
    \includegraphics[width=0.48\textwidth]{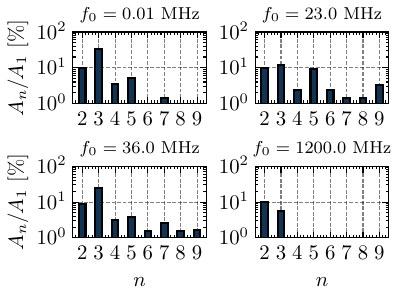}
    \caption{\textbf{Frequency-Domain Spectrum.} The Fast Fourier Transform (FFT) spectra of the output current $I(t)$, where amplitudes of the higher-order harmonics $A_n$ at $n \cdot f_0$ are normalized to the amplitude of the fundamental driving frequency $A_1$ at $f_0$. The spectra show a transition from a simple nonlinearity, where most energy is located at small orders $n$, to a more complex nonlinearity close to the cutoff frequency $f_c \approx 26~\mathrm{MHz}$. At large frequencies ($f_0 = 1200~\mathrm{MHz}$) the current loses its nonlinearity due to linear capacitive shunting.}
    \label{fig:example_fft_spectra}
\end{figure}

\begin{figure*}[t]
    \centering
    \begin{subfigure}[b]{0.49\textwidth}
        \centering
        \includegraphics[height=0.22\textheight]{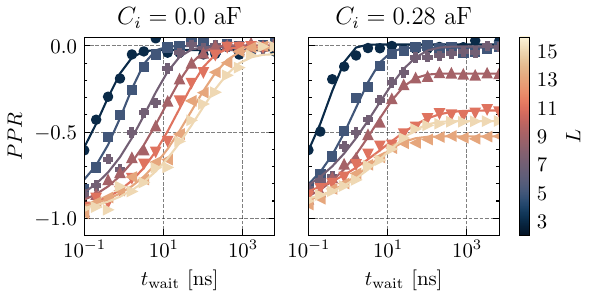}
        \caption{Paired-Pulse Relaxation}
        \label{fig:PPR_vs_L}
    \end{subfigure}
    \hfill
    \begin{subfigure}[b]{0.49\textwidth}
        \centering
        \includegraphics[height=0.2\textheight]{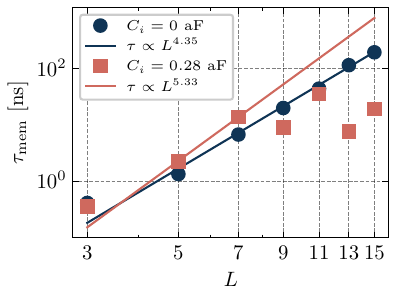}
        \caption{Fading Memory Time Constant}
        \label{fig:tau_mem_vs_L}
    \end{subfigure}
    \caption{\textbf{Fading and non-volatile Memory Regime.} \textbf{(a)} Paired Pulse Ratio (PPR) representing the network's memory as a function of the waiting time $t_\textrm{wait}$ between successive input pulses (forgetting curves). At $t_\textrm{wait} = 0.1~\mathrm{ns}$ the network's state is still fully affected by the first pulse when the second pulse arrives. As waiting time increases the network progressively forgets the initial pulse ($PPR \to 0$) in case of the isolated system ($C_i = 0$). For a screened system ($C_i = 0.28$), small networks ($L \le 5$) exhibit complete fading memory when $t_\text{wait}$ increases as well. However, networks much larger than the screening length $\Lambda = 3$ trap charges in the bulk, resulting in a non-zero asymptote indicative of persistent memory. \textbf{(b)} The characteristic memory timescale $\tau_\mathrm{mem}$, extracted from a stretched exponential fit of the PPR forgetting curves increases with lattice length $L$. This increment is unbounded for isolated system ($C_i = 0$) which does not couple to a beneath substrate. For screened system at coupling $C_i = 0.28~\mathrm{aF}$, the fading memory scaling is bounded by the screening length, resulting in a plateau marking the non-volatile memory regime.}
    \label{fig:nonlinear_and_memory}
\end{figure*}

\begin{figure}[b]
    \centering
    \includegraphics[width=0.48\textwidth]{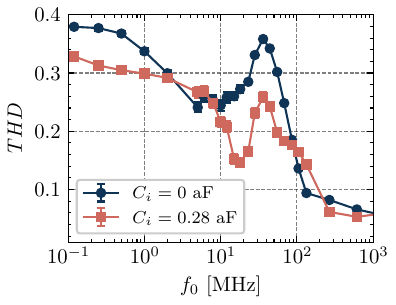}
    \caption{\textbf{Harmonic Distortion} Total Harmonic Distortion (THD) of the output current $I(t)$ as a function of the driving frequency $f_0$. The THD transitions from nonlinear resistive to linear displacement currents when increasing the frequency $f_0$. A peak in nonlinearity occurs directly at the cutoff frequency $f_c \approx 26~\mathrm{MHz}$.}
    \label{fig:THD_vs_L}
\end{figure}

When comparing an isolated network of substrate coupling $C_i = 0$ and a screened network of substrate coupling $C_i = 0.28~\mathrm{aF}$, we observe two crucial differences. The isolated device yields a screening length of $\Lambda \to \infty$. This infinite length, allows capacitive interactions to couple the input and output globally (see FIG.~\ref{fig:pot_dist_along_cycle_zero}). As there are no localized capacitive sinks to ground needed to be filled, the 2D lattice reduces electrostatically into a single lumped equivalent capacitor. As at large frequency, only these capacitive interactions contribute to the electric current, the admittance $Y_I$ approaches a size-independent asymptote. In the screened environment, charges pay an electric field tax at each nanoparticle to ground resulting in the high frequency limit becoming size dependent. Especially, as the screening length is finite at $\Lambda = 3$ for $C_i = 0.28~\mathrm{aF}$, charges must tunnel through the device reaching the output (see FIG.~\ref{fig:pot_dist_along_cycle_028}). The amount of required tunnelling events is system size dependent and lead to a delay, marking a negative phase shift between output response and input signal. Computationally, this transforms the network from a derivative filter into an integrator capable of accumulating signal history.~\cite{mohamed2015modeling,kamble2018coexistence,romero2020meminductor,salaoru2014coexistence}

In linear RC circuit theory, the frequency dependent admittance of a RC element is defined to be $Y(\omega) = G + i\omega C$ with conductance $G = 1/R$ and capacitance $C$. The point at which resistive and capacitive contributions are equal yields the cutoff frequency i.e. $\omega_{RC} = \frac{G}{C} = \frac{1}{RC} = \frac{1}{\tau_{RC}}$ with time constant $\tau_{RC} = RC$. To estimate the cutoff frequency marking the crossover point from nonlinear resistive tunneling to linear capacitive shunting, we measure the frequency $f_c$ at which $I^\textrm{tun}$ and $I^\textrm{disp}$ intersect. FIG.~\ref{fig:bandwidth} compares the cutoff frequency for different system sizes $L$ and electrostatic substrate couplings $C_i$. Given the fact $\mathbf{C}_{U}$ only consists of mutual nanoparticle to electrode interactions $C_{i\alpha}$, displacement currents (equation~\eqref{eq:I_disp}) are not explicitly dependent on $C_i$. As a result, the cutoff frequency is independent of the oxide thickness $d_\textrm{ox}$. The scaling of $f_c \propto L^{-2.3}$ establishes a bandwidth when operating nanoparticle networks. Accordingly an input voltage of $U_0(t) = \tilde{U}_0\sin(2 \pi f_c t)$ ensures the network is operated in a regime at which both nonlinear components by tunneling and phase-shifting characteristics by displacement contribute. Extrapolating the scaling in FIG.~\ref{fig:bandwidth} towards $L=1$ yields a value of order $f_c(1) \approx 4.0~\mathrm{GHz}$. This value compares to the theoretical value from linear RC theory $f_{RC} = \frac{1}{2\pi\tau_{RC}} \approx 2.8~\mathrm{GHz}$.

\section{Nonlinearity and Memory}\label{sec:nonlinearity_memory}

\begin{figure*}[t]
    \centering
    \includegraphics[width=\textwidth]{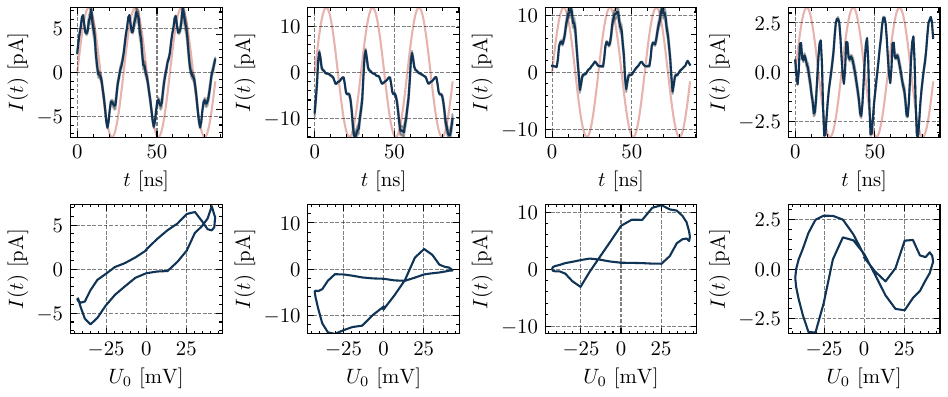}
    \caption{\textbf{Control-tuned Network Responses.} Time-domain and phase space plots of the total output current $I(t)$ given a sinusoidal input voltage $U_0(t) = \tilde{U}_0 \sin(2\pi f_c t$ and fixed but randomly chosen control electrode voltages. Here, $f_c$ denotes the cutoff frequency. Depending on the given set of control electrode voltages, the sinusoidal input signal is either heavily distorted with high-order harmonic contributions (4th column), changed into a pulse-like responses (2nd or 3rd column) or passed more or less unaffected (1st column).}
    \label{fig:examples_responses}
\end{figure*}

The nanoparticle network must simultaneously map input signals into a higher-dimensional nonlinear space and retain a temporal memory of past inputs to be utilized for complex dynamic tasks. We quantify the network's nonlinear capacity using the Total Harmonic Distortion (THD) equation~\eqref{eq:THD} of the output electric current $I(t)$ under a sinusoidal input $U_0(t) = \tilde{U}_0\sin(2 \pi f_0 t)$.

Calculations of THD are based on frequency spectra such as displayed in FIG.~\ref{fig:example_fft_spectra}. FIG.~\ref{fig:THD_vs_L} shows the THD dependence on the input frequency $f_0$ for a lattice of size $L = 9$. The regime of small frequencies is dominated by nonlinear resistive currents, maximizing THD as the Coulomb blockade onset produces powerful higher harmonics. As $f_0$ approaches the network's cutoff frequency $f_c \approx 26~\mathrm{MHz}$, the THD exhibits a distinct peak. For large frequencies, the THD collapses. This high-frequency roll-off is the consequence of the capacitive shunting regime. The linear displacement current $I^\textrm{disp}(t)$ completely overtakes the tunneling current $I^\textrm{tun}(t)$ and the signal transformation inevitably linearizes.

To quantify the network's memory, we calculate the Paired-Pulse Ratio (PPR) defined in equation~\eqref{eq:PPR} as a function of the inter-pulse waiting time $t_\text{wait}$. Each curve represents a forgetting curve. At small $t_\text{wait}$, the second response is reduced given the network is still affected from the first pulse. When we increase $t_\text{wait}$, charges eventually tunnel back to the electrodes and the network forgets the first pulse before the second pulse arrives. However, there is a crucial system size dependent effect.

For a small network ($L \le 7$) whose length is comparable to the electrostatic screening length ($\Lambda \approx 3$), the network bulk is entirely penetrated by the electrode's voltage. This electrostatic proximity of bulk and electrodes ensures that injected charges always escape during the zero-bias phase, yielding a PPR that eventually approaches zero. Notably, in the ideal limit of isolated networks ($\Lambda \to \infty$, where $C_i = 0~\mathrm{aF}$), this complete relaxation is universally achieved regardless of lattice size $L$, as the absence of distributed capacitive traps prevents bulk charge retention entirely. As the lattice becomes larger ($L \ge 9$) in the screened environment ($\Lambda \approx 3$), we observe a crucial transition in the relaxation dynamics. For larger networks, charges injected deep into the bulk during the initial pulse lack the interaction length to tunnel back to the boundaries. Charges become trapped, which manifests as a permanent suppression of the subsequent pulse response, freezing the PPR at a non-zero negative asymptote. This transition shifts the network from exhibiting fading short-term memory to more persistent, non-volatile-like memory. A microscopic visualization of the non-volatile regime is observed in the potential landscape snapshot in FIG.~\ref{app:non_volatile_memory_example}.

To quantify the fading memory horizon, we model the temporal relaxation of the PPR using a stretched exponential function. FIG.~\ref{fig:tau_mem_vs_L} plots the resulting memory time constant $\tau_\textrm{mem}$ from the fit as a function of the lattice size $L$. The data reveals that the temporal depth of the fading memory scales ($\tau \propto L^{4.3}$ or $\tau \propto L^{5.3}$) with network length $L$, providing an essential tuning parameter for matching the network's memory to the temporal length of a given target task. However, it must be noted that this scaling strictly bounds to finite length scales. As $L \gg \Lambda$, the network undergoes the transition into the non-volatile regime, and fading memory is permanently arrested.

These results establish the design constraints for dynamic applications. Operating below the cutoff frequency ($f_0 \ll f_c$) produces a strong nonlinear response but no memory, whereas operating above the cutoff frequency ($f_0 \gg f_c$) suppresses the nonlinear response. Consequently, the system size $L$ should be selected to obtain an appropriate cutoff frequency while achieving the desired trade-off between nonlinearity and memory. For a given $L$, the oxide thickness $d_\textrm{ox}$ can then be adjusted to tune the transition from the fading to the non-volatile memory regime.

\section{Dynamic Tuneability}

While the previous sections established that the nanoparticle network possesses nonlinearity and either fading or non-volatile-like memory, we must now quantify the absolute degree of control we can exert over its dynamic response (expressivity). To evaluate the expressivity of the network we quantify its effective volume defined in equation~\eqref{eq:eff_volume}. For this we varied the control electrode voltages 1000 times for different lattice sizes $L$ and substrate couplings $C_i$. The input electrode voltages is set to $U_0(t) = \tilde{U}_0\sin(2 \pi f_c t)$ with cutoff frequency $f_c$ based on FIG.~\ref{fig:bandwidth}. FIG.~\ref{fig:examples_responses} shows time domain and phase space plots for four selected control voltage combinations. Depending on the external bias, the sinusoidal input is either heavily distorted with high-order harmonic contributions, changed into a pulse-like response or passed more or less unaffected.

\begin{figure}[tb]
    \centering
    \includegraphics[width=0.48\textwidth]{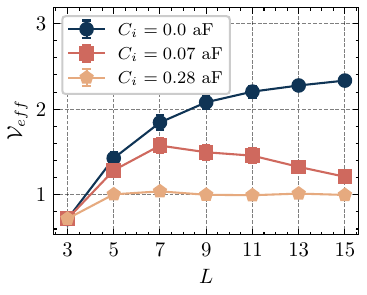}
    \caption{\textbf{Effective Volume.} The effective volume $\mathcal{V}_{\text{eff}}$ scales with system length $L$. The isolated network ($C_i = 0.0~\mathrm{aF}$, $\Lambda \to \infty$) achieves a larger absolute volume. For screened networks ($C_i \neq 0.0~\mathrm{aF}$), the increase in effective volume $\mathcal{V}_{\text{eff}}$ with system size is bounded as control electrode voltages cannot reach the network's bulk given the finite screening length $\Lambda$.}
    \label{fig:V_eff_vs_ss}
\end{figure}

FIG.~\ref{fig:V_eff_vs_ss} shows the system-size-dependent effective volume $\mathcal{V}_\textrm{eff}$ for three substrate couplings ($C_i = 0.0~\mathrm{aF}$, $\Lambda \to \infty$; $C_i = 0.07~\mathrm{aF}$, $\Lambda \approx 6$; and $C_i = 0.28~\mathrm{aF}$, $\Lambda \approx 3$). The effective volume is scaled so that it yields a value of $V_\textrm{eff} = 1$ for a network of size $L = 9$ and substrate coupling $C_i = 0.28~\mathrm{aF}$. The error bars include the stochastic nature of the KMC algorithm as well as the sampling error given the finite number of control electrode voltage configurations (see section~\ref{sec:bootstrapping_error_propagation}). As the physical system size $L$ increases, the effective volume $\mathcal{V}_\textrm{eff}$ generally grows. Larger networks contain more independently tunable junctions, providing a richer substrate for control-induced manipulation. In general, less screened networks attain larger effective volumes at a given system size because their larger electrostatic screening length allows static control fields applied at the boundaries to penetrate deeper into the bulk, thereby modulating a larger fraction of the internal nanoparticles. For the isolated network ($C_i = 0$, $\Lambda \to \infty$), the increase in effective volume is unbounded because all nanoparticles can participate in the nonlinear mixing processes regardless of system size. In contrast, for screened networks with substrate couplings $C_i \in {0.07, 0.28}~\mathrm{aF}$ and screening lengths $\Lambda \in {6, 3}$, the effective volume eventually saturates. This saturation arises because the network develops an electrostatically opaque bulk once the system size exceeds the screening length. In this regime, control fields originating at the perimeter decay before reaching the network center. Consequently, the deep bulk becomes electrostatically isolated and no longer contributes to the nonlinear signal-processing dynamics.

\begin{figure}[b]
    \centering
    \includegraphics[width=0.48\textwidth]{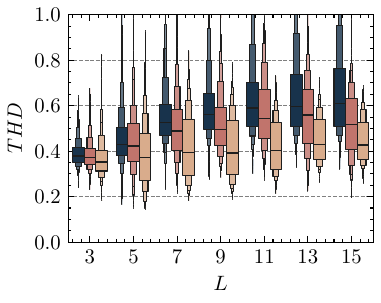}
    \caption{\textbf{Total Harmonic Distortion.} Probability distributions of the Total Harmonic Distortion (THD) across the set of random control voltage configurations. At optimal length, the network acts as an expressive, voltage-controlled nonlinear filter at large variance. Increasing the system size above the screening length $\Lambda$, truncates the dynamic range, confirming that maximum tunability requires a balance between screening length $\Lambda$ and system size $L$.}
    \label{fig:thd_dist_vs_ss}
\end{figure}

A similar topological constraint is also reflected in the probability distributions of the Total Harmonic Distortion (THD) across all control electrode combinations presented in FIG.~\ref{fig:thd_dist_vs_ss}. The vertical spread of these distributions represents the dynamic range over which the control electrodes can tune the electric current nonlinearity. At optimal system size the control electrodes stretch the THD to larger nonlinear values. In this regime, the device functions as an expressive and programmable nonlinear filter.

The findings in FIG.~\ref{fig:V_eff_vs_ss} and FIG.~\ref{fig:thd_dist_vs_ss} establish an experimental design rule. Increasing the system size is advantageous only if the electrostatic substrate screening ($\Lambda$) is co-engineered, based on the oxide thickness $d_\text{ox}$, to match the system length $L$. While maximizing device tuneability generally requires minimizing screening by isolating the system further from the substrate, this approach presents drawbacks. Specifically, as demonstrated in section~\ref{sec:ac_response_bandwidth} and section~\ref{sec:nonlinearity_memory}, these less-screened configurations lack both a negative phase shift and non-volatile-like behavior. Consequently, optimizing the device ultimately requires a trade-off tailored to the specific application.

\section{\label{sec:disorder}Structural Disorder}

\begin{figure}[t]
    \centering
    \includegraphics[width=0.48\textwidth]{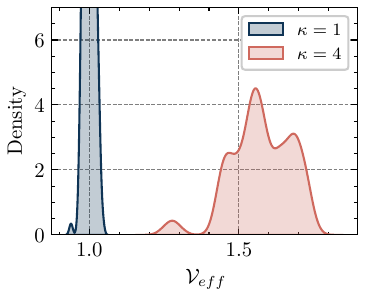}
    \caption{\textbf{Expansion of the Response Space via Symmetry Breaking.} Effective volume ($\mathcal{V}_{\text{eff}}$) distributions for a uniform network ($\kappa = 1$) and a 32-configuration disordered ensemble ($\kappa = 4$). The uniform baseline is normalized to $\mathcal{V}_{\text{eff}} = 1$. The disordered networks exhibit a universal increase in effective volume, confirming spatial heterogeneity as a robust enhancer of device expressivity. Resampling is applied to account for KMC and finite-sampling errors.}
    \label{fig:V_eff_dists_vs_kappa}
\end{figure}

Throughout the preceding sections, the network was treated as a uniform lattice featuring identical junction resistances ($R_1 = 25~\mathrm{M\Omega}$). Experimentally, structural disorder can be introduced via a second Type-2 molecular junction, which disrupts internal spatial symmetries. This disorder is quantified utilizing the bimodal distribution detailed in equation~\eqref{eq:bimonal_dis_dist}. We subsequently compare the uniform baseline network ($\kappa=1$) with a disordered topology ($\kappa=4$). To isolate the universal characteristics of this disorder from individual spatial variations, an ensemble of 32 randomly generated configurations is simulated for the $\kappa=4$ network. In a uniform network, the current distribution is dictated entirely by the electrostatic potential landscape, governed by control electrode voltages and the screening length $\Lambda$. Conversely, within disordered devices, high-resistance Type-2 molecules act as internal tunneling bottlenecks, adding resistive complexity onto the inherent electrostatic complexity.

For this analysis, we selected an exemplary uniform network with a size of $L = 9$ and a substrate coupling of $C_i = 0.28~\mathrm{aF}$, applying these same baseline parameters to our $\kappa = 4$ networks. The effective volume is calculated using equation~\eqref{eq:eff_volume}. Consistent with FIG.~\ref{fig:V_eff_vs_ss}, we scale the results so that the uniform reference yields a mean volume of $\mathcal{V}_\textrm{eff} = 1$. As demonstrated in FIG.~\ref{fig:V_eff_dists_vs_kappa}, introducing this symmetry breaking expands the response space, increasing the effective volume of the disordered network ($\kappa = 4$) relative to the uniform baseline ($\kappa = 1$). To account for both the KMC error and the finite size of the control voltage sample, these distributions incorporate the resampling technique detailed in section~\ref{sec:bootstrapping_error_propagation}. Furthermore, because the $\kappa = 4$ distribution spans the full ensemble of 32 resistance configurations, the data reveals that device tuneability improves regardless of the specific spatial layout. For experimental design, this establishes spatial heterogeneity as a robust enhancer of tuneability yield.

\begin{figure}[b]
    \centering
    \includegraphics[width=0.48\textwidth]{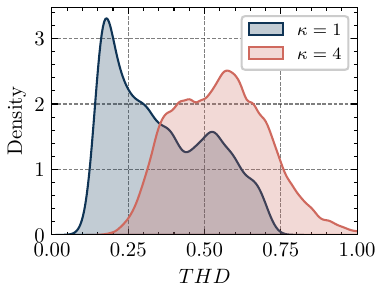}
    \caption{\textbf{Total Harmonic Distortion (THD) Distributions for uniform and disordered Networks.} THD evaluated across sampled control voltages for the $\kappa = 1$ baseline and the $\kappa = 4$ ensemble. The introduction of structural disorder shifts the response toward higher harmonic energy, with some configurations reaching $THD = 1$. Although the distributions partially overlap, the uniform network remains constrained to a less expressive response compared to its spatially heterogeneous counterparts.}
    \label{fig:THD_vs_disorder}
\end{figure}

Analogous to FIG.~\ref{fig:thd_dist_vs_ss}, we also evaluate the distribution of Total Harmonic Distortion (THD) values (equation~\eqref{eq:THD}) across the sampled control electrode voltages. FIG.~\ref{fig:THD_vs_disorder} compares the uniform $\kappa = 1$ baseline against the ensemble of 32 disordered $\kappa = 4$ networks. The disordered distribution noticeably shifts toward larger THD values. In fact, some realizations reach $THD = 1$, indicating that the device response contains as much energy in its higher harmonics as in the fundamental amplitude.

While the introduction of structural disorder ($\kappa = 4$) yields a universal increase in the effective volume $\mathcal{V}_{\text{eff}}$ compared to the uniform baseline ($\kappa = 1$), the THD distributions display a degree of overlap. This discrepancy highlights the fundamental difference between the scalar THD magnitude and high-dimensional expressivity dictated by $\mathcal{V}_\textrm{eff}$. THD is a one-dimensional projection that quantifies the total harmonic energy, compressing both phase information and the distribution of energy across individual harmonics. A symmetric $\kappa = 1$ network can occasionally be tuned into configurations with high total harmonic energy, resulting in the observed THD overlap. However, the uniform network remains constrained to a lower-dimensional manifold within the $2(M-1)$-dimensional response space due to its spatial symmetry. In contrast, the spatial heterogeneity of the $\kappa = 4$ topology breaks these internal symmetries, allowing the control voltages to independently manipulate specific harmonic amplitudes and their relative phases ($\Delta \phi_m$). Consequently, even if a specific disordered configuration exhibits a marginally lower THD, its capacity to navigate a larger set of independent computational states renders it more expressive, as captured by the superior $\mathcal{V}_{\text{eff}}$.
  
\section{Conclusion}

In this work, we have demonstrated the potential of metallic nanoparticle networks as an expressive substrate for physical computing. By replacing the traditional digital readout phase in Reservoir Computing with static control electrodes, we established a neuromorphic architecture that routes one-dimensional voltage-encoded input signals through high-dimensional dynamical regimes.

We first analyzed the system intrinsically, with all control electrodes turned off, revealing a fundamental trade-off between nonlinearity and phase-shift memory governed by the cutoff frequency. At low frequencies, the response is dominated by nonlinear charge tunneling, whereas linear charge displacement dictates the high-frequency regime. Operating at this cutoff frequency yields an optimal balance between nonlinear transformation at high harmonic orders and phase-delay memory. Crucially, the type of memory is defined by the thickness of the underlying oxide layer, which dictates the nanoparticle-to-substrate coupling and the resulting electrostatic screening length. Networks on a thick oxide layer (isolated networks) possess an infinite screening length, allowing charges to always escape. This provides a fading memory characteristic. Conversely, networks on a thinner oxide layer exhibit a finite screening length. If the network length exceeds this screening length, charges become tightly trapped inside the network bulk, transitioning the system into a persistent, non-volatile regime.

Next, we activated the static control electrodes to evaluate the system's practical expressivity. We observed that while the expressivity of isolated networks scales unbounded with system size, screened networks only benefit from spatial scaling if the system is not significantly larger than the screening length. Finally, to fundamentally enhance the device's expressivity, we introduced structural disorder. By breaking internal spatial symmetries, heterogeneous junctions allow the control voltages to independently manipulate specific harmonic amplitudes and their relative phases. 

Together, these results establish operating frequency, electrostatic screening length, and junction heterogeneity as three experimentally accessible design parameters for engineering nanoparticle networks as tunable physical computing substrates. Future experiments can now leverage this framework to tailor devices for specific temporal tasks, ultimately benchmarking these multi-electrode architectures against standard models like time-dependent XOR and NARMA.

\begin{acknowledgments}

The author(s) declare financial support was received for the research, authorship, and/or publication of this article. This work was funded by the Deutsche Forschungsgemeinschaft (DFG, German Research Foundation) through project 433682494–SFB 1459.

\end{acknowledgments}

\bibliography{td_np_nets}

\clearpage
\onecolumngrid
\setcounter{equation}{0}
\setcounter{figure}{0}
\setcounter{table}{0}
\setcounter{page}{1}
\setcounter{section}{0}
\renewcommand{\theequation}{S\arabic{equation}}
\renewcommand{\thefigure}{S\arabic{figure}}
\renewcommand{\thetable}{S\arabic{table}}
\renewcommand{\thesection}{S\arabic{section}}

\begin{figure*}[tb]
    \centering
    \begin{subfigure}[b]{0.49\textwidth}
    \centering
    \includegraphics[height=0.25\textheight]{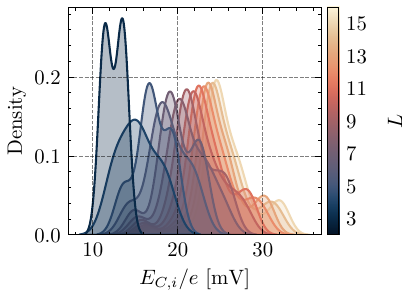}
    \caption{$C_i = 0$}
    \label{app:charging_energies_ci=0}
    \end{subfigure}
    \hfill
    \begin{subfigure}[b]{0.49\textwidth}
    \centering
    \includegraphics[height=0.25\textheight]{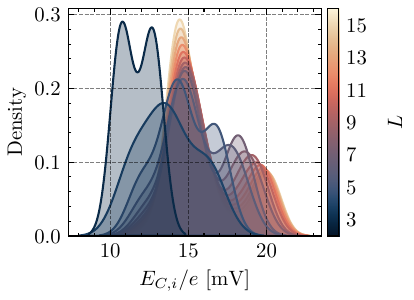}
    \caption{$C_i = 0.28~\mathrm{aF}$}
    \label{app:charging_energies_ci=028}
    \end{subfigure}
    \caption{\textbf{System Size Dependence of local Voltage Thresholds associated to the Single-Electron Charging Energy.}\textbf{(a)} When increasing the lattice length $L$ without coupling each nanoparticle to a substrate ($C_i = 0$), the distribution of voltages shifts to larger values as added charges must polarize the entire nanoparticle network. \textbf{(b)} Introducing a substrate coupling of $C_i = 0.28~\mathrm{aF}$ bounds the increase of voltage thresholds to a fixed distribution at large values of $L$. A finite electrostatic interaction dictated by a screening of $\Lambda = 3$ only polarizes close neighbors as portions of the electric field are terminated at the substrate.}
    \label{app:charging_energies}
\end{figure*}

\section{Charging Energy}

Adding a discrete electron to a specific nanoparticle $i$ requires the local electrostatic potential to overcome the Coulomb energy $E_{C,i}$, as defined in equation~\eqref{eq:e_energy_single}. This charging energy corresponds to a voltage threshold $E_{C,i}/e$, which depends on both the topological coordinate of the nanoparticle $i$ and its capacitance to the substrate, $C_i$.

FIG.~\ref{app:charging_energies} compares the distributions of these local voltage thresholds across lattices of varying sizes ($L \times L$) for substrate couplings of $C_i \in \{0.0, 0.28\}~\mathrm{aF}$. In an isolated electrostatic environment where $C_i = 0$ (yielding an infinite screening length, $\Lambda \to \infty$), charging a single nanoparticle forces the entire lattice to polarize. As the lattice length $L$ increases, the energy cost to charge a nanoparticle grows indefinitely, causing the threshold distribution to shift unbounded toward higher energies (see FIG.~\ref{app:charging_energies_ci=0}). Conversely, in a screened system with a substrate coupling of $C_i = 0.28~\mathrm{aF}$ and a screening length of $\Lambda = 3$, the substrate absorbs a portion of the electric field lines. Because this confines the polarization to a finite radius, the local Coulomb penalty becomes independent of the system size $L$ once the charged nanoparticle is surrounded by a sufficient shell of neighbors. Consequently, the distributions shown in FIG.~\ref{app:charging_energies_ci=028} converge toward a limiting distribution in large lattices.

To ensure a fair comparison across all topological dimensions, we use the maximum charging energy $\max_i\{E_{C,i}/e\}$ summarized in TABLE~\ref{tab:voltage_scaling}, as a voltage scaling parameter. This approach guarantees that variations in external electrode voltages consistently and properly excite the system's nonlinear single-electron dynamics.

\begin{table}[htbp]
\caption{\label{tab:voltage_scaling} Voltage Thresholds [mV] based on the maximum charging energies $\max_i\{E_{C,i}/e\}$ for different lattice length $L \times L$ across three substrate couplings $C_i$ [aF].}
\begin{ruledtabular}
\begin{tabular}{lccccccc}
$C_i$ \textbackslash{} $L$ & 3 & 5 & 7 & 9 & 11 & 13 & 15 \\
\colrule
0.0  & 13.58 & 19.34 & 22.94 & 26.14 & 28.59 & 30.67 & 32.52 \\
0.07 & 13.37 & 18.58 & 21.45 & 23.53 & 24.89 & 25.68 & 26.36 \\
0.28 & 12.77 & 16.80 & 18.56 & 19.50 & 20.07 & 20.29 & 20.45 \\
\end{tabular}
\end{ruledtabular}
\end{table}

\section{FDM-based Electrostatics}

\begin{figure*}[t]
\centering
\begin{subfigure}[b]{0.49\textwidth}
\centering
\includegraphics[height=0.25\textheight]{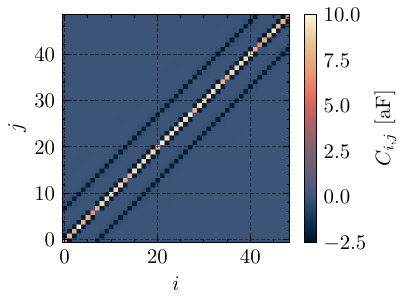}
\caption{FDM-based Capacitance Matrix}
\label{app:capacitance_matrix_FMD}
\end{subfigure}
\hfill
\begin{subfigure}[b]{0.49\textwidth}
\centering
\includegraphics[height=0.25\textheight]{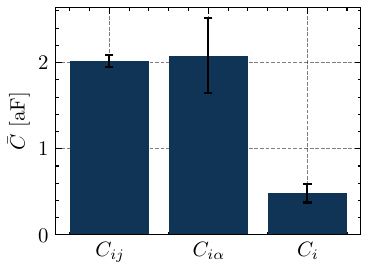}
\caption{Electrostatic Hierarchy}
\label{app:capacitance_comparsion}
\end{subfigure}
\caption{\textbf{FDM-based Electrostatics.} \textbf{(a)} The $7 \times 7$ capacitance matrix $\mathbf{C}$ is defined by nearest neighbor interaction with next-neighbors approaching $C_{ij} = 0$ exhibiting a banded structure. Higher-order neighboring couplings are suppressed. \textbf{(b)} Solving the Poisson equation for a $7 \times 7$ lattice of nanoparticles using a 3D Finite Difference Method yields a physically accurate electrostatic hierarchy. The electrostatic interaction are dominated by nearest neighbor couplings $C_{ij}$ and $C_{i\alpha}$ between nanoparticles $i$ and $j$ or nanoparticle $i$ and electrode $\alpha$. Conversely, the substrate coupling $C_i$ is screened by the dense network.}
\label{app:capacitance_FMD}
\end{figure*}

A $7 \times 7$ lattice was constructed using geometric parameters representative of the experimental device described in~\cite{bose2015evolution}. We use a nanoparticle radius of $10~\mathrm{nm}$, an inter-particle gap of $1~\mathrm{nm}$, and a uniform $\text{SiO}_2$ dielectric thickness of $d_\text{ox} = 35~\mathrm{nm}$. The substrate is assumed to be a conducting ground plane located beneath the oxide layer.

Using a 3D Finite Difference Method (FDM) to solve the Poisson equation results in the full capacitance matrix $\mathbf{C}$ shown in FIG.~\ref{app:capacitance_matrix_FMD}. This matrix is heavily banded, given the first-order off-diagonal elements representing immediate horizontal or vertical nearest neighbors dominate the coupling. The suppression of second-nearest neighbors via electrostatic screening justifies truncating the mutual capacitance coupling to exclusively nearest neighbors in our approximation. Taking the average of each interaction type derives the electrostatic hierarchy in FIG.~\ref{app:capacitance_comparsion}. The network is dominated by the horizontal interactions $C_{ij}$ between two nanoparticles $i$ and $j$, and $C_{i\alpha}$ between nanoparticle $i$ and electrode $\alpha$. The average nanoparticle to substrate interaction $C_{i}$ is much smaller.

\begin{figure*}[tbh]
    \centering
    \includegraphics[width=\textwidth]{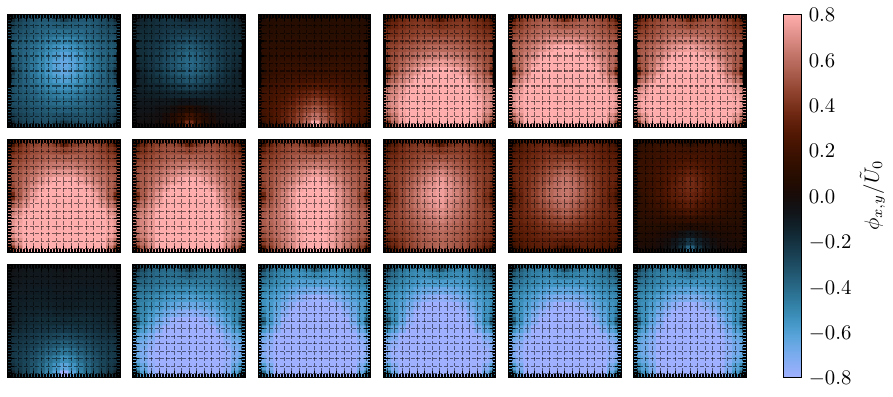}
    \caption{\textbf{Potential Distribution along a Cycle of the Input sinusoidal Signal for $C_i = 0$.} Given a lattice of $L = 15$ nanoparticles, the potential value $\phi_{x,y}$ at lattice position $(x,y)$ is shown during a cycle of the input signal $U_0(t) = \tilde{U}_0\sin(2\pi f_0 t)$ scaled by the input's amplitude $\tilde{U}_0$. Given the screening length of $\Lambda \to \infty$, the network switches as a whole from positive to negative potentials when ever the input changes its polarity.}
    \label{fig:pot_dist_along_cycle_zero}
\end{figure*}

\begin{figure*}[tbh]
    \centering
    \includegraphics[width=\textwidth]{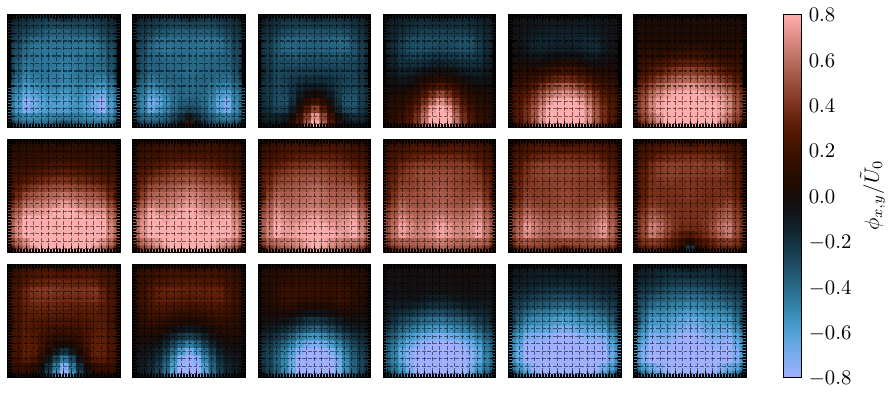}
    \caption{\textbf{Potential Distribution along a Cycle of the Input sinusoidal Signal for $C_i = 0.28~\mathrm{aF}$.} Given a lattice of $L = 15$ nanoparticles, the potential value $\phi_{x,y}$ at lattice position $(x,y)$ is shown during a cycle of the input signal $U_0(t) = \tilde{U}_0\sin(2\pi f_0 t)$ scaled by the input's amplitude $\tilde{U}_0$. Given the screening length of $\Lambda = 3$, the potential density propagates trough the network as the input changes its polarity.}
    \label{fig:pot_dist_along_cycle_028}
\end{figure*}

Assuming nearest neighbor interaction for nanoparticles of same size in a one-dimensional lattice ($L \times 1$), all mutual capacitance values are identical ($C_{ij} = C_m$) and all substrate couplings are uniform ($C_{i} = C_s$). For a particle $i$, the total capacitance then yields $[\mathbf{C}]_{ii} = C_g + 2C_m$, while the capacitance matrix off-diagonal elements are $[\mathbf{C}]_{i,i\pm1} = -C_m$. Using equation~\eqref{eq:pot_vec}, the total charge of this particle is
\begin{equation}\label{app:total_charge_1D}
    q_i = (C_g + 2C_m)\phi_i - C_m\phi_{i-1} - C_m\phi_{i+1}.
\end{equation}
For the pure screening profile, we consider a charge-free situation, allowing us to set equation~\eqref{app:total_charge_1D} to zero, leading to the discrete Laplace equation
\begin{equation}\label{app:d_lap}
    C_m(\phi_{i-1} - 2\phi_i + \phi_{i+1}) - C_g\phi_i = 0
\end{equation}
and in its continuum limit to the 1D Poisson equation
\begin{equation}\label{app:c_lap}
    \frac{\partial^2\phi}{\partial x^2} - \frac{C_g}{C_m a^2} = 0.
\end{equation}
where $a$ is the lattice constant. This equation yields an exponential decay of $\phi(x) \propto \exp( -x/\Lambda )$, where the characteristic screening length is defined in units of the lattice constant as
\begin{equation}\label{app:soliton_Length}
    \Lambda = \sqrt{\frac{C_m}{C_g}}.
\end{equation}
FIG.~\ref{fig:pot_dist_along_cycle_zero} visually proofs that a substrate coupling of $C_i = 0$ yields a screening length of $\Lambda \to \infty$. When changing the polarity of an input curve $U_0(t) = \tilde{U}_0\sin(2\pi f_0 t)$, the network's potential landscape reacts as a whole and switches its polarity immediately as well. When introducing a substrate coupling of $C_i = 0.28~\mathrm{aF}$, the screening length is finite at $\Lambda = 3$. Accordingly in FIG.~\ref{fig:pot_dist_along_cycle_028}, charges must tunnel trough the device and change the potential landscape step by step, starting at the bottom where the sine is applied. 

\section{Singular Value Spectrum of the Electrode-to-Nanoparticle Coupling}

The element $M_{i\alpha}$ of the electrode-to-nanoparticle coupling matrix $\mathbf{M} = \mathbf{C}^{-1}\mathbf{C}_U \in \mathbb{R}^{N_p \times N_e}$ dictates how the voltage at electrode $\alpha$ capacitively induces a potential on nanoparticle $i$ (equation.~\eqref{eq:pot_vec}). Physically, $\mathbf{M}$ acts as a linear operator that maps the $N_e$-dimensional space of external electrode voltages $\vec{U}(t)$ onto the $N_p$-dimensional space of nanoparticle potentials $\vec{\phi}(t)$. To assess the effective control dimensionality provided by the external electrodes, we perform a Singular Value Decomposition (SVD) of the coupling matrix $\mathbf{M} = \mathbf{W} \mathbf{\Sigma} \mathbf{V}^T$, where $\mathbf{W}$ and $\mathbf{V}$ are orthogonal matrices containing the spatial potential modes and electrode voltage modes, respectively, and $\mathbf{\Sigma} = \mathrm{diag}(\sigma_1, \sigma_2, \dots, \sigma_{N_e})$ contains the singular values ordered by magnitude ($\sigma_1 \ge \sigma_2 \ge \dots \ge \sigma_{N_e} > 0$). Qualitatively, each singular value $\sigma_k$ quantifies the strength with which an independent, orthogonal combination of electrode voltages (a column of $\mathbf{V}$) can independently actuate a corresponding potential pattern across the nanoparticle network (a column of $\mathbf{W}$). 

\begin{figure}[tbh]
    \centering
    \includegraphics[width=0.5\textwidth]{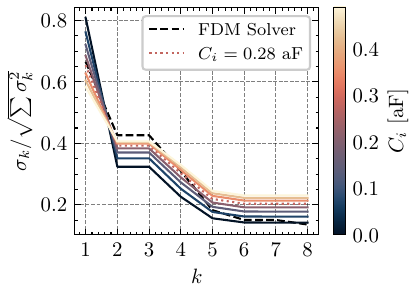}
    \caption{\textbf{Control Dimensionality.} Normalized singular values ($\sigma_k / \sqrt{\sum \sigma_k^2}$) of the electrode-to-nanoparticle coupling matrix $\mathbf{M} = \mathbf{C}^{-1}\mathbf{C}_U$. Both the exact FDM solver (dashed black line) and the capacitance model across different substrate couplings $C_i$ maintain a strictly non-vanishing spectrum. This confirms that all $N_e = 8$ control electrode voltages linearly project onto independent orthogonal potential modes across the network.}
    \label{fig:control_dimension}
\end{figure}

FIG.~\ref{fig:control_dimension} shows the normalized singular value spectrum $\sigma_k / \sqrt{\sum_j \sigma_j^2}$ for $N_e = 8$ control electrodes across different levels of substrate couplings $C_i$, alongside the exact finite-difference method (FDM) solution. For all considered substrate couplings, the spectrum remains strictly non-vanishing ($\sigma_k > 0$ for all $k$). This confirms that the coupling matrix retains full column rank, ensuring that no control electrodes are redundant and that the external inputs span an 8-dimensional subspace of linearly independent potential drives across the network.

\section{Temperature Dependence}\label{app:temperature_limit}

The nonlinear transport characteristics of nanoparticle networks rely on the discrete quantization of charge and the energy penalty of the Coulomb blockade. These phenomena are sensitive to temperature. As the thermal energy $k_B T$ approaches the characteristic charging energy $E_{C,i}$ defined in equation~\eqref{eq:e_energy_single}, the thresholds for tunneling are progressively smeared by the exponential nature of the tunneling rate equation~\eqref{eq:tunneling}.

To quantify this quantum-to-classical transition, we evaluated the transport characteristics of a $9 \times 9$ nanoparticle network at substrate coupling $C_i = 0.28~\mathrm{aF}$ across a wide temperature range. We measure the constant steady state output electrode electric current $I$ to a constant input electrode bias $U_0$.
\begin{figure*}[t]
    \centering
    \begin{subfigure}[b]{0.48\textwidth}
    \centering
    \includegraphics[height=0.24\textheight]{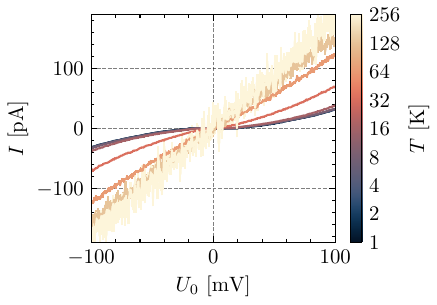}
    \caption{Thermal Smearing of $I-U$ Curves}
    \label{fig:iv_curves_temperature}
    \end{subfigure}
    \hfill
    \begin{subfigure}[b]{0.48\textwidth}
    \centering
    \includegraphics[height=0.24\textheight]{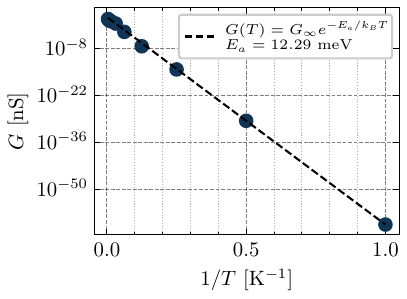}
    \caption{Arrhenius Conductivity Scaling}
    \label{fig:arrhenius_plot}
    \end{subfigure}
    \caption[Thermal Degradation of the Coulomb Gap.]{\textbf{Thermal Degradation of the Coulomb Gap.} \textbf{(a)} Drain current $I$ to input voltage $U_0$ dependence in a $9 \times 9$ network. At $1~\mathrm{K}$, a nonlinear zero-current Coulomb gap persists. As temperature increases, thermal fluctuations smear the blockade thresholds. At $256~\mathrm{K}$, transport becomes purely linear, converging to the resistor network solution (black dashed line). \textbf{(b)} Arrhenius plot of the zero-bias conductance $G$ versus inverse temperature $1/T$. The fit across the complete range (black dashed line) captures the thermally activated tunneling regime. The extracted global activation energy is $E_a = 12.29~\mathrm{meV}$.}
    \label{fig:network_iv_and_arrhenius}
\end{figure*}
As shown in Fig.~\ref{fig:iv_curves_temperature}, the $1~\mathrm{K}$ $I-U$ curve exhibits a zero current Coulomb gap. Because $k_B T \ll E_{C,i}$, tunneling events that would increase the electrostatic free energy of the system are suppressed. However, as the network's temperature is increased, thermal fluctuations provide the necessary activation energy to overcome these Coulomb penalties. This smearing of the discrete charging thresholds results in a narrowing of the Coulomb blockade gap. The $I-U$ curves become softer at intermediate temperatures ($4~\mathrm{K}$, $16~\mathrm{K}$, $32~\mathrm{K}$), and a thermally activated sub-threshold current emerges near $0~\mathrm{V}$. This transition underlines the physical necessity of operating the network in small temperature ranges to ensure the network is dominated by nonlinearities rather than stochastic thermal noise.

At $256~\mathrm{K}$, the available thermal energy of $k_B T \approx 22~\mathrm{meV}$ exceeds the network's maximum characteristic charging energy of $\max_i\{E_{C,i}\} \approx 19.5~\mathrm{meV}$. In this high-temperature limit, the discrete tunnel junctions behave collectively as an Ohmic resistor grid approaching a linear current to voltage dependence.

To extract the activation energy $E_a$ of the network, we analyze the thermal scaling of the $U_0 = 0$ conductance $G$. Because transport in the Coulomb blockade regime is thermally activated, the zero-bias conductance must follow the Arrhenius law $G(T) = G_\infty \exp(-E_a / k_B T)$. FIG.~\ref{fig:arrhenius_plot} presents the Arrhenius plot of the conductance $G$ versus inverse temperature $1/T$. The fit yields an activation energy of $E_a = 12.29~\mathrm{meV}$. This value is lower than the absolute maximum charging energy of the network ($\max_i\{E_{C,i}\} \approx 19.5~\mathrm{meV}$). This is a consequence of the fact that transport does not require charges to overcome the single highest electrostatic barrier. Instead, $E_a$ characterizes the energy required to excite a single charge that can percolate through the least-resistive junction of the network.

Here, $E_a$ defines the absolute thermal boundary for the network's operational nonlinear regime. We need to at least ensure that the network is operated at a temperature which is about an order of magnitude smaller than the temperature relating to $E_a/k_B \approx 142~\mathrm{K}$. The probability of spontaneous thermal tunneling is $\mathcal{P} \propto \exp(-E_a/k_B T)$. A temperature of $142~\mathrm{K}$ yields a probability of $\mathcal{P} = e^{-1} \approx 37\%$ making thermal fluctuations still the dominant factor. Reducing the temperature to $T \sim 30~\mathrm{K}$ at probability $\mathcal{P} = e^{-5} \approx 0.5\%$ or even $T \sim 14.2~\mathrm{K}$ at probability $\mathcal{P} = e^{-10} \approx 0.004\%$ allows Coulomb blockade driven transport be the dominant factor. For simplicity, throughout this work we set the temperature to $T = 0.1~\mathrm{K}$ ensuring thermally activated tunnelling is negligible.

\section{Kinetic Monte Carlo Model}\label{app:kmc}

The network's dynamics are simulated using a continuous-time Kinetic Monte Carlo (KMC) algorithm.~\cite{mensing2024kinetic,wasshuber2001computational,gillespie1977exact} For a system occupying a discrete integer charge state at time $t$ with a corresponding local potential vector $\vec{\phi}$ and fixed electrode voltages $\vec{U}$, a single KMC iteration proceeds as follows:

\begin{enumerate}
    \item \textbf{Rate Evaluation:} The discrete tunneling rates $\Gamma_k$ of equation~\eqref{eq:tunneling} for all possible single-electron transitions (inter-particle and electrode-to-particle) are evaluated based on the potentials $\vec{\phi}$ using the established free-energy equation~\eqref{eq:free_energy}.
    \item \textbf{Stochastic Time Advance:} The time $\Delta t$ until the next tunneling event occurs is a stochastic variable sampled from an exponential distribution governed by the total system exit rate, $\Gamma_{\text{total}} = \sum \Gamma_k$:
    \begin{equation}\label{eq:dwell_time}
        \Delta t = -\frac{\ln(r_1)}{\Gamma_{\text{total}}}
    \end{equation}
    where $r_1 \in (0, 1]$ is a uniform random number.
    \item \textbf{Event Selection:} The specific transition $m$ is selected via inverse transform sampling on the cumulative rate vector. We employ a binary search to find the index $m$ satisfying $\Gamma_{\text{cum}, m} \ge r_2 \cdot \Gamma_{\text{total}}$ for a second uniform random number $r_2$, thereby reducing the algorithmic complexity of event selection to $\mathcal{O}(\log N_{\text{events}})$.
    \item \textbf{Linear Potential Update:} Rather than executing a computationally expensive $\mathcal{O}(N_p^2)$ matrix-vector multiplication as in equation~\eqref{eq:pot_vec} to update the system state after a tunneling event, we exploit the linearity of the electrostatic equations. The potential vector is updated incrementally in $\mathcal{O}(N_p)$ time. For an internal transition of an electron from nanoparticle $i$ to nanoparticle $j$, the new potential is simply:
    \begin{equation}
        \vec{\phi}_{\text{new}} = \vec{\phi}_{\text{old}} + e \left( [\mathbf{C}^{-1}]_{*, j} - [\mathbf{C}^{-1}]_{*, i} \right)
    \end{equation}
    where $[\mathbf{C}^{-1}]_{*, k}$ denotes the $k$-th column vector of the inverse capacitance matrix.
\end{enumerate}

Standard KMC implementations assume that transition rates remain constant between discrete events. To accurately simulate the dynamics driven by the continuous input $U_0(t)$, we use a time-slicing approach.~\cite{anderson2007modified,vestergaard2015temporal} The integration period is discretized into intervals $\Delta t_{\text{sim}}$, chosen to satisfy the sampling theorem for the highest relevant harmonic of the drive frequency $f_0$. Within each slice $[t_k, t_k + \Delta t_{\text{sim}})$, the external voltages $\vec{U}$ are held constant, and the standard KMC loop evolves the system. If the clock advances beyond the current slice boundary ($t + \Delta t > t_k + \Delta t_{\text{sim}}$), the KMC loop pauses, the pending event is discarded, and the internal clock is exactly synchronized to the boundary. At the boundary, $\vec{U}(t)$ is updated, the induced offset charges $\mathbf{C}_U \vec{U}(t)$ are analytically added to $\vec{\phi}$, and the stochastic generation resumes with newly evaluated rates.

Because the system is stochastically driven, extracting observables requires ensemble averaging across $N$ independent trajectory realizations, denoted by $\langle \dots \rangle$. The total current $\langle I_{\alpha}^{\text{total}}(t_k) \rangle$ measured at external electrode $\alpha$ is the superposition of the resistive single-electron tunneling current and the capacitive displacement current.

The tunneling current $\langle I_{\alpha}^{\text{tun}}(t_k) \rangle$ is calculated directly from the ensemble-averaged net transition rates across the electrode junction at time $t_k$
\begin{equation}\label{eq:I_tun}
    \langle I_{\alpha}^\text{tun}(t_k) \rangle = \frac{e}{N} \sum_{m=1}^N \left[ \Gamma_{i \to \alpha, m}(t_k) - \Gamma_{\alpha \to i, m}(t_k)\right].
\end{equation}
The displacement current, driven by the continuous polarization of the network, is derived from the time-derivative of the local potentials. This current is computed analytically during post-processing using the smooth, ensemble-averaged potential vector $\langle \vec{\phi}(t_k) \rangle$:
\begin{equation}\label{eq:I_disp}
    \langle I_{\alpha}^{\text{disp}}(t_k) \rangle = \sum_{i=1}^{N_p} [\mathbf{C}_U]_{i\alpha} \frac{d}{dt} \left( \langle \phi_i(t_k) \rangle - U_\alpha(t_k) \right)
\end{equation}
By combining these components, we resolve the dynamical response, separating the nonlinear sequential tunneling from the linear capacitive shunting. For notational simplicity, the ensemble average brackets $\langle \dots \rangle$ are implied for all reported continuous-time observables.

\section{Bootstrapping and Error Propagation}\label{sec:bootstrapping_error_propagation}

When transforming the total current $I(t)$ into performance metrics such as Total Harmonic Distortion (equation~\eqref{eq:THD}) we must propagate the error $\sigma_I(t)$ across each current trajectory into the performance metric. The current is statistically distributed as a Gaussian $I \sim \mathcal{N}(I, \sigma_I^2)$, justified by the Central Limit Theorem. Besides, when calculating the effective volume (equation~\eqref{eq:eff_volume}) we measure a finite amount of currents defined by the number of control electrode voltage configurations ($N_{\text{cfg}}$). This induces an additional sampling bias. To simultaneously account for both sources of uncertainty we use a resampling algorithm. \paragraph{Bootstrapping:} When calculating the effective volume, we first draw a new synthetic dataset of size $N_{\text{cfg}}$ with replacement. \paragraph{Parametric Resampling:} For each configuration selected in the draw, its specific currents are not taken as static values. Instead, an ensemble of synthetic current values is generated by independently drawing from their respective Gaussian distributions $I \sim \mathcal{N}(I, \sigma_I^2)$. As we are handling time-dependent current curves $I(t_k)$, the current is resampled at each time step $t_k$ given the local standard deviation $\sigma_I(t_k)$. \paragraph{Metric Evaluation:} The target metric $y = f(I)$, is computed over this entirely synthetic ensemble. By repeating this resampling procedure $M$ times, we construct an ensemble of metric values $\{y^{(1)}, y^{(2)}, \dots, y^{(M)}\}$ representing the full probability density function. The final performance metric is defined as the arithmetic mean of this ensemble $\langle y \rangle = \frac{1}{M} \sum y^{(m)}$. The propagated uncertainty (error bounds) is defined by the standard deviation of this distribution $\sigma_y$.

\section{Microscopic Origin of Non-Volatile Memory}

\begin{figure*}[t]
\centering
\begin{subfigure}[b]{0.48\textwidth}
\centering
\includegraphics[height=0.25\textheight]{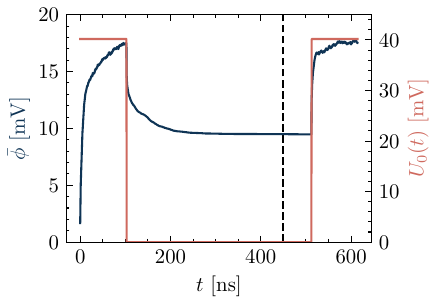}
\caption{Time-Domain Potential}
\label{app:pot_vs_t_example}
\end{subfigure}
\hfill
\begin{subfigure}[b]{0.48\textwidth}
\centering
\includegraphics[height=0.25\textheight]{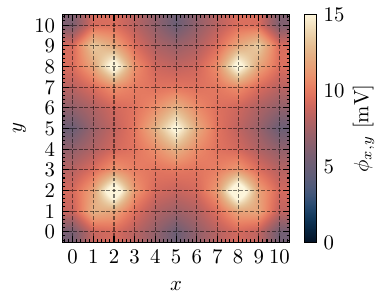}
\caption{Potential Landscape at Zero-Bias}
\label{app:non_volatile_memory_example}
\end{subfigure}
\caption{\textbf{Internal temporal and spatial Signatures of Charge Trapping.} \textbf{(a)} Time-domain evolution of the spatial averaged nanoparticle potentials during a paired-pulse step input sequence. During the zero-bias waiting period, the potentials fail to relax to the electrostatic ground state, indicating the presence of trapped charges. \textbf{(b)} A 2D spatial heatmap of the nanoparticle potentials across the network lattice, captured at the end of the waiting period (vertical dashed line in FIG a). The map shows that charges are permanently trapped within the network bulk. This frozen spatial configuration is the microscopic origin of the non-volatile-like memory measured in the paired-pulse response.}
\label{app:non_volatile_memory_micro}
\end{figure*}

To verify the mechanism driving the transition from fading short-term memory to non-volatile-like memory observed in large networks ($L \ge 9$), here we examine the microscopic state of a network of size $L=11$ and substrate coupling $C_i=0.28~\mathrm{aF}$. In section~\ref{sec:nonlinearity_memory}, we postulated that the non-zero asymptote of the Paired Pulse Ratio (PPR) arises from charges being trapped within the bulk of the network, given the finite screening length of $\Lambda = 3$.

FIG.~\ref{app:pot_vs_t_example} tracks the temporal evolution of the averaged nanoparticle potential during the write-wait-write sequence. During the initial write pulse, the high external bias overcomes the Coulomb blockade, injecting charges and increasing the potentials. Crucially, when the driving voltage is removed ($t_\text{wait}$), the potentials do not relax back to the zero-bias ground state ($0~\mathrm{V}$). Instead, they plateau at a non-zero state. Because the network length $L=11$ exceed the electrostatic screening length of $\Lambda = 3$, the electrostatic pull of the grounded electrodes is negligible deep within the bulk.

FIG.~\ref{app:non_volatile_memory_example} visualizes this frozen microscopic state. By extracting a 2D spatial heatmap of the nanoparticle potentials at the end of the zero-bias waiting period indicated by the vertical dashed line in FIG.~\ref{app:pot_vs_t_example}, we can directly observe the potential distribution. The heatmap reveals localized high potential islands isolated within the center of the network. These are trapped charges. This trapped spatial configuration permanently changes the electrostatic landscape. When the second pulse is applied, it encounters a fundamentally different electrostatic starting state than the first pulse, leading to a suppressed current response and the resulting non-volatile asymptote.

\end{document}